\RequirePackage{lineno}
\documentclass[aps,prd,twocolumn,showpacs,superscriptaddress,groupedaddress]{revtex4}  
\usepackage{graphicx}
\usepackage{epsfig}
\usepackage{dcolumn}
\usepackage{bm}
\usepackage[usenames]{color} 
\usepackage{subfigure}
\usepackage{amsmath}
\usepackage{amssymb}
\usepackage{float}
\usepackage[colorlinks=true, pdfstartview=FitV, linkcolor=blue, citecolor=blue, urlcolor=blue]{hyperref}

\newcommand{\beq}{\begin{equation}}
\newcommand{\eeq}{\end{equation}}
\newcommand{\barr}{\begin{eqnarray}}
\newcommand{\earr}{\end{eqnarray}}

\def\bq{\begin{quote}}
\def\eq{\end{quote}}

\def\spose#1{\hbox to 0pt{#1\hss}}
\def\lsim{\mathrel{\spose{\lower 3pt\hbox{$\mathchar"218$}}
 \raise 2.0pt\hbox{$\mathchar"13C$}}}
\def\gsim{\mathrel{\spose{\lower 3pt\hbox{$\mathchar"218$}}
 \raise 2.0pt\hbox{$\mathchar"13E$}}}

\def\bs{${B_s^0}$}

\def\bsdec{${B_s^0 \rightarrow J/\psi \phi}$}
\def\bsf{$B_s^0 \rightarrow J/\psi f_2^{\prime}(1525)$}
\def\bskk{$B_s^0 \rightarrow J/\psi K^+ K^-$}

\def\bddec{${B_d^0 \rightarrow J/\psi K^*}$}

\def\D0{D\O }

\def\GeVp{ {\ifmmode \;{{\mbox{\mathrm GeV}} / {\mbox\mathrm c}} \else
${{\mbox{\mathrm GeV}} / {\mbox\mathrm c}}$ \fi }}
\def\MeVp{ {\ifmmode \;{{\mbox{\mathrm MeV}} / {\mbox\mathrm c}} \else
${{\mbox{\mathrm MeV}} / {\mbox\mathrm c}}$ \fi }}
\def\MeV{ {\ifmmode \;{{\mbox{\mathrm MeV}} / {\mbox\mathrm c}^2} \else
${{\mbox{\mathrm MeV}} / {\mbox\mathrm c}^2}$ \fi }}
\def\GeV{ {\ifmmode \;{{\mbox{\mathrm GeV}} / {\mbox\mathrm c}^2} \else
${{\mbox{\mathrm GeV}} / {\mbox\mathrm c}^2}$ \fi }}

\begin{document}

\hspace{5.2in} 
\mbox{FERMILAB-PUB-12-104-E}

\title{Study of the decay  {\boldmath $B_s^0 \rightarrow J/\psi f_2^{\prime}(1525)$} 
in  {\boldmath  $\mu^+ \mu^- K^+K^-$} final states}

%
\affiliation{LAFEX, Centro Brasileiro de Pesquisas F\'{i}sicas, Rio de Janeiro, Brazil}
\affiliation{Universidade do Estado do Rio de Janeiro, Rio de Janeiro, Brazil}
\affiliation{Universidade Federal do ABC, Santo Andr\'e, Brazil}
\affiliation{University of Science and Technology of China, Hefei, People's Republic of China}
\affiliation{Universidad de los Andes, Bogot\'a, Colombia}
\affiliation{Charles University, Faculty of Mathematics and Physics, Center for Particle Physics, Prague, Czech Republic}
\affiliation{Czech Technical University in Prague, Prague, Czech Republic}
\affiliation{Center for Particle Physics, Institute of Physics, Academy of Sciences of the Czech Republic, Prague, Czech Republic}
\affiliation{Universidad San Francisco de Quito, Quito, Ecuador}
\affiliation{LPC, Universit\'e Blaise Pascal, CNRS/IN2P3, Clermont, France}
\affiliation{LPSC, Universit\'e Joseph Fourier Grenoble 1, CNRS/IN2P3, Institut National Polytechnique de Grenoble, Grenoble, France}
\affiliation{CPPM, Aix-Marseille Universit\'e, CNRS/IN2P3, Marseille, France}
\affiliation{LAL, Universit\'e Paris-Sud, CNRS/IN2P3, Orsay, France}
\affiliation{LPNHE, Universit\'es Paris VI and VII, CNRS/IN2P3, Paris, France}
\affiliation{CEA, Irfu, SPP, Saclay, France}
\affiliation{IPHC, Universit\'e de Strasbourg, CNRS/IN2P3, Strasbourg, France}
\affiliation{IPNL, Universit\'e Lyon 1, CNRS/IN2P3, Villeurbanne, France and Universit\'e de Lyon, Lyon, France}
\affiliation{III. Physikalisches Institut A, RWTH Aachen University, Aachen, Germany}
\affiliation{Physikalisches Institut, Universit\"at Freiburg, Freiburg, Germany}
\affiliation{II. Physikalisches Institut, Georg-August-Universit\"at G\"ottingen, G\"ottingen, Germany}
\affiliation{Institut f\"ur Physik, Universit\"at Mainz, Mainz, Germany}
\affiliation{Ludwig-Maximilians-Universit\"at M\"unchen, M\"unchen, Germany}
\affiliation{Fachbereich Physik, Bergische Universit\"at Wuppertal, Wuppertal, Germany}
\affiliation{Panjab University, Chandigarh, India}
\affiliation{Delhi University, Delhi, India}
\affiliation{Tata Institute of Fundamental Research, Mumbai, India}
\affiliation{University College Dublin, Dublin, Ireland}
\affiliation{Korea Detector Laboratory, Korea University, Seoul, Korea}
\affiliation{CINVESTAV, Mexico City, Mexico}
\affiliation{Nikhef, Science Park, Amsterdam, the Netherlands}
\affiliation{Radboud University Nijmegen, Nijmegen, the Netherlands}
\affiliation{Joint Institute for Nuclear Research, Dubna, Russia}
\affiliation{Institute for Theoretical and Experimental Physics, Moscow, Russia}
\affiliation{Moscow State University, Moscow, Russia}
\affiliation{Institute for High Energy Physics, Protvino, Russia}
\affiliation{Petersburg Nuclear Physics Institute, St. Petersburg, Russia}
\affiliation{Instituci\'{o} Catalana de Recerca i Estudis Avan\c{c}ats (ICREA) and Institut de F\'{i}sica d'Altes Energies (IFAE), Barcelona, Spain}
\affiliation{Uppsala University, Uppsala, Sweden}
\affiliation{Lancaster University, Lancaster LA1 4YB, United Kingdom}
\affiliation{Imperial College London, London SW7 2AZ, United Kingdom}
\affiliation{The University of Manchester, Manchester M13 9PL, United Kingdom}
\affiliation{University of Arizona, Tucson, Arizona 85721, USA}
\affiliation{University of California Riverside, Riverside, California 92521, USA}
\affiliation{Florida State University, Tallahassee, Florida 32306, USA}
\affiliation{Fermi National Accelerator Laboratory, Batavia, Illinois 60510, USA}
\affiliation{University of Illinois at Chicago, Chicago, Illinois 60607, USA}
\affiliation{Northern Illinois University, DeKalb, Illinois 60115, USA}
\affiliation{Northwestern University, Evanston, Illinois 60208, USA}
\affiliation{Indiana University, Bloomington, Indiana 47405, USA}
\affiliation{Purdue University Calumet, Hammond, Indiana 46323, USA}
\affiliation{University of Notre Dame, Notre Dame, Indiana 46556, USA}
\affiliation{Iowa State University, Ames, Iowa 50011, USA}
\affiliation{University of Kansas, Lawrence, Kansas 66045, USA}
\affiliation{Kansas State University, Manhattan, Kansas 66506, USA}
\affiliation{Louisiana Tech University, Ruston, Louisiana 71272, USA}
\affiliation{Boston University, Boston, Massachusetts 02215, USA}
\affiliation{Northeastern University, Boston, Massachusetts 02115, USA}
\affiliation{University of Michigan, Ann Arbor, Michigan 48109, USA}
\affiliation{Michigan State University, East Lansing, Michigan 48824, USA}
\affiliation{University of Mississippi, University, Mississippi 38677, USA}
\affiliation{University of Nebraska, Lincoln, Nebraska 68588, USA}
\affiliation{Rutgers University, Piscataway, New Jersey 08855, USA}
\affiliation{Princeton University, Princeton, New Jersey 08544, USA}
\affiliation{State University of New York, Buffalo, New York 14260, USA}
\affiliation{Columbia University, New York, New York 10027, USA}
\affiliation{University of Rochester, Rochester, New York 14627, USA}
\affiliation{State University of New York, Stony Brook, New York 11794, USA}
\affiliation{Brookhaven National Laboratory, Upton, New York 11973, USA}
\affiliation{Langston University, Langston, Oklahoma 73050, USA}
\affiliation{University of Oklahoma, Norman, Oklahoma 73019, USA}
\affiliation{Oklahoma State University, Stillwater, Oklahoma 74078, USA}
\affiliation{Brown University, Providence, Rhode Island 02912, USA}
\affiliation{University of Texas, Arlington, Texas 76019, USA}
\affiliation{Southern Methodist University, Dallas, Texas 75275, USA}
\affiliation{Rice University, Houston, Texas 77005, USA}
\affiliation{University of Virginia, Charlottesville, Virginia 22901, USA}
\affiliation{University of Washington, Seattle, Washington 98195, USA}
\author{V.M.~Abazov} \affiliation{Joint Institute for Nuclear Research, Dubna, Russia}
\author{B.~Abbott} \affiliation{University of Oklahoma, Norman, Oklahoma 73019, USA}
\author{B.S.~Acharya} \affiliation{Tata Institute of Fundamental Research, Mumbai, India}
\author{M.~Adams} \affiliation{University of Illinois at Chicago, Chicago, Illinois 60607, USA}
\author{T.~Adams} \affiliation{Florida State University, Tallahassee, Florida 32306, USA}
\author{G.D.~Alexeev} \affiliation{Joint Institute for Nuclear Research, Dubna, Russia}
\author{G.~Alkhazov} \affiliation{Petersburg Nuclear Physics Institute, St. Petersburg, Russia}
\author{A.~Alton$^{a}$} \affiliation{University of Michigan, Ann Arbor, Michigan 48109, USA}
\author{G.~Alverson} \affiliation{Northeastern University, Boston, Massachusetts 02115, USA}
\author{M.~Aoki} \affiliation{Fermi National Accelerator Laboratory, Batavia, Illinois 60510, USA}
% ...
%
%
\collaboration{The D0 Collaboration\footnote{Visitors from
$^{a}$Augustana College, Sioux Falls, SD, USA,
$^{b}$The University of Liverpool, Liverpool, UK,
$^{c}$UPIITA-IPN, Mexico City, Mexico,
$^{d}$DESY, Hamburg, Germany,
,
$^{e}$SLAC, Menlo Park, CA, USA,
$^{f}$University College London, London, UK,
$^{g}$Centro de Investigacion en Computacion - IPN, Mexico City, Mexico,
$^{h}$ECFM, Universidad Autonoma de Sinaloa, Culiac\'an, Mexico
and
$^{i}$Universidade Estadual Paulista, S\~ao Paulo, Brazil.
$^{\ddag}$Deceased.
}} \noaffiliation
\vskip 0.25cm
\date{\today}

\begin{abstract}
We investigate the decay  $B_s^0 \rightarrow J/\psi K^+ K^-$ for invariant masses of the $K^+K^-$ pair 
in the range $1.35<M(K^+K^-)<2$ GeV. The data sample corresponds 
to an integrated luminosity of 10.4 fb$^{-1}$ of $p \overline p$ collisions at $\sqrt{s} = 1.96$ TeV
accumulated with the D0 detector at the Fermilab Tevatron collider.
From the study of the invariant mass and spin of the $K^+K^-$ system, we find evidence 
for the two-body decay \bsf\ and measure the  relative branching fraction
 of the decays \bsf\ and \bsdec\ to be
$R_{f_2^{\prime}/\phi} = 0.19 \pm 0.05 \rm \thinspace{(stat)} \pm 0.04 \rm \thinspace{(syst)}$. 

\end{abstract}


\maketitle


\section{\label{sec:intro}Introduction}

In the standard model~\cite{PDG}, the mixing of quarks
originates from their interactions with the Higgs field causing
the quark mass eigenstates to be different from the quark flavor
eigenstates. Constraints on the mixing phases are obtained from measurements of the
decays of neutral mesons. 
Decays to final states that are common to both partners of a 
neutral-meson doublet are of particular importance. 
The interference between the
amplitude of the direct decay and the amplitude of
the decay following the particle-antiparticle oscillation
 may lead to a $CP$-violating
asymmetry between decays of mesons and antimesons.
The decays $B_s^0 \rightarrow J/\psi X$, where $X$ stands for a
pair of charged kaons or pions, are a  sensitive probe for 
new phenomena because the $CP$-violating phase that appears in
such decays is predicted in the standard model to be 
close to zero with high precision~\cite{smpsis}. 

 The first observation of the decay  sequence $B_{s}^{0} \rightarrow J/\psi f_{2}^{\prime}(1525)$,
$f_{2}^{\prime}(1525) \rightarrow K^+K^-$,  was recently reported by LHCb~\cite{lhcbf2}. 
The spin $J=2$ assignment for the $K^+K^-$ pair was based 
on excluding pure $J=0$. In this article,  we confirm the presence of
 the decay $B_s^0 \rightarrow J/\psi K^+K^-$
with $K^+K^-$ invariant masses $M(K^+K^-)$ close to 1.52 GeV and  we determine that the spin of
 the $K^+K^-$ resonance is consistent with $J=2$ and is preferred over $J=0$ or $J=1$ assignments.
We identify the resonance as the  $f_2^{\prime}(1525)$ meson and 
measure the branching fraction relative to the well established decay  \bsdec.

\section{\label{sec:detector}Detector}

The D0 detector consists of a central tracking system, a calorimetry system and
muon detectors, as detailed in Refs.~\cite{run2det,layer0,muon}. The central
tracking system comprises  a silicon microstrip tracker (SMT) and a central
fiber tracker, both located inside a 1.9~T superconducting solenoidal
magnet.  The tracking system is designed to optimize tracking and vertexing
for pseudorapidities $|\eta|<3$, where  $\eta = -\ln[\tan(\theta/2)]$ and  $\theta$ is the 
polar angle with respect to the proton beam direction.
   
The SMT can reconstruct the $p\overline{p}$ interaction vertex (PV) 
for interactions with at least three tracks with a precision
of 40 $\mu$m in the plane transverse to the beam direction and determine
the impact parameter of any track relative to the PV with a precision between
20 and 50~$\mu$m, depending on the number of hits in the SMT.
    
The muon detector surrounds the calorimeter.
It consists of a central muon system covering the pseudorapidity
region $|\eta|<1$ and a forward muon system covering the pseudorapidity region $1<|\eta|<2$.
Both  systems consist of a layer of drift tubes and scintillators inside
1.8~T toroidal magnets and two similar layers outside the toroids.

\section{\label{sec:event} Event Reconstruction and Candidate Selection}

The analysis presented here is based on a data sample corresponding to
an integrated luminosity of 10.4 fb$^{-1}$ accumulated
between February 2002 and September 2011 at the Fermilab Tevatron collider.
Events are collected with single-muon and dimuon triggers. 
Some triggers require the presence of tracks with a large impact parameter  with respect
to the  PV. The events selected
exclusively by these triggers are removed from our sample.

Candidate $B^0_s \rightarrow J/\psi K^+K^-$, $J/\psi \rightarrow \mu^+ \mu^-$ events
are required to include two oppositely charged muons accompanied by two oppositely charged tracks.
Both muons are required to be detected in the muon chambers inside the toroidal magnet,
and at least one of the muons is required to be also detected outside the toroid.
Each of the four tracks is required to have at least one SMT hit. In addition, the kaon candidates
are required to have at least two hits in the SMT, at least two hits
in the central fiber tracker,
 and a total of at least eight hits in both detectors.

To form $B_s^0$ candidates, muon pairs in the invariant mass range
$2.9< M(\mu^+ \mu^-) < 3.3$  GeV, consistent with $J/\psi$ decay,
are combined with pairs of oppositely charged particles (assigned the kaon mass), 
consistent with production at a common vertex. A  kinematic fit under the
$B_s^0$ decay hypothesis  constrains $M(\mu^+ \mu^-)$
to the Particle Data Group~\cite{PDG} value of the $J/\psi$ mass and  the four tracks  to a common vertex.
The trajectories of the four $B_s^0$ decay products are
adjusted according to this  kinematic fit.
In events where multiple candidates satisfy these requirements, we select
the candidate with the best  fit probability.
The  $\chi^2$ of the fit is required to be less than 15, for a total
number of degrees of freedom of 4.
We require that the transverse momenta of the $B_{s}^0$ 
and $K^{\pm}$ mesons are larger than 8 and 0.7 GeV, respectively.
To suppress background from the decay 
$B^0 \rightarrow J/\psi K^*(892)$, we require the kaon
pair to have an invariant mass greater than 1 GeV under
the $K^{\pm}\pi^{\mp}$ hypothesis. 

To reconstruct the PV, we select tracks that 
do not originate from the candidate $B_s^0$ decay 
and apply a constraint to the average beam-spot position~\cite{run2det} in the transverse plane.
We define the signed  decay length of a \bs\ meson, $L^B_{xy}$, 
as the vector pointing from the PV to the decay vertex, projected on the
\bs\  transverse momentum $p_T$.
The proper decay time of a \bs\ candidate is given by
 $t = M_{B_s}\vec L_{xy}^B \cdot \vec{p}/(p_T^2)$
where $M_{B_s}$ is the world-average \bs\ mass~\cite{PDG},
and $\vec p$ is the particle momentum.
To increase \bs\ signal purity and reject prompt background, we require 
the proper decay length to be greater than $200$  $\mu$m.
The distribution of the uncertainty on the proper decay length peaks around
$25$ $\mu$m and has a long tail extending to several hundred microns.
To remove poorly reconstructed events in the tail, we require the
uncertainty of the proper decay length to be less than $100$  $\mu$m.

\section{\label{sec:mc} Monte Carlo Samples}

Some aspects of this  analysis require information 
that cannot be derived from data. We rely on Monte Carlo (MC)
simulated samples to derive templates of the distributions of the
signal \bsf\ and of the main background components, and to derive
the detector acceptance as a function of decay angles, and
 the relative acceptance for the decays \bsdec\ and \bsf.  

We use  {\sc pythia}  \cite{pythia} to generate $B_{s}^{0}$ mesons 
and {\sc evtgen}~\cite{evtgen} to simulate their decay. 
In all simulated samples the final states are assumed to have no polarization.
The samples have been processed by the  detector simulation  and
the standard  event reconstruction.
To take into account the effects of the instantaneous
luminosity, the MC samples are overlaid with data events collected
 during random beam crossings.
We have generated events containing the decays \bsf, \bsdec,
$B^0 \rightarrow J/\psi K_2^{*}(1430)$, and $B^0 \rightarrow J/\psi K_0^{*}(1430)$.

\section{\label{sec:signal} $B_s^0 \rightarrow J/\psi K^+K^-$ signal extraction}
 
The distribution of the $B^0_s$ candidate mass $M(J/\psi K^+K^-)$ as a function
of  $M(K^+K^-)$ for accepted events is shown in
Fig.~\ref{fig:kkvsbs_data}.
The data show a structure consistent with the decay  \bsf.
Another possibility for the observed structure
near $M(K^+K^-)=1.5$~GeV is the decay $B^0_s \rightarrow J/\psi f_{0}(1500)$.
The two states have very different branching fractions to $\pi^+ \pi^-$ and $K^+ K^-$. 
The branching fractions~\cite{PDG} are
${\cal B} (f_0(1500)$$\rightarrow$$ \pi \pi)$=($34.9 \pm 2.3)$\%, 
${\cal B} (f_0(1500)$$\rightarrow$$K\overline{K})$=$(8.6\pm1.0)$\%,
 ${\cal B} (f_2^{\prime}(1525)$$ \rightarrow \pi \pi)$=$(0.82 \pm
 0.15)$\%, and ${\cal B} (f_2^{\prime}(1525)$$\rightarrow K\overline{K}$=$(88.7 \pm 2.2)$\%.
If the $M(K^+K^-)$  peak is due to the $f_0(1500)$, a larger
peak is expected near  $M(\pi^+\pi^-)=1.5$~GeV.
If the $M(K^+K^-)$  peak is due to the  $f_2^{\prime}(1525)$ meson, a negligibly 
small peak is expected in the $M(\pi^+\pi^-)$  distribution.
From the  lack of significant \bs\ signal in the $\pi^+\pi^-$ channel,
presented in  Fig.~\ref{fig:pipivsbs}, we conclude that
the observed state is not the $f_0(1500)$ meson.

However, as observed by others~\cite{lhcbf2,babarz}, a similar distribution may also result from decays of
$B^0$ mesons where the $J/\psi$ meson is accompanied by a  $K^*$ resonance, and a pion
from the decay $K^* \rightarrow K^{\pm} \pi^{\mp}$ is assigned the kaon mass.


\begin{widetext}

\begin{figure}[H]
\begin{center}
\includegraphics*[height=18.0cm,width=1.0\textwidth]{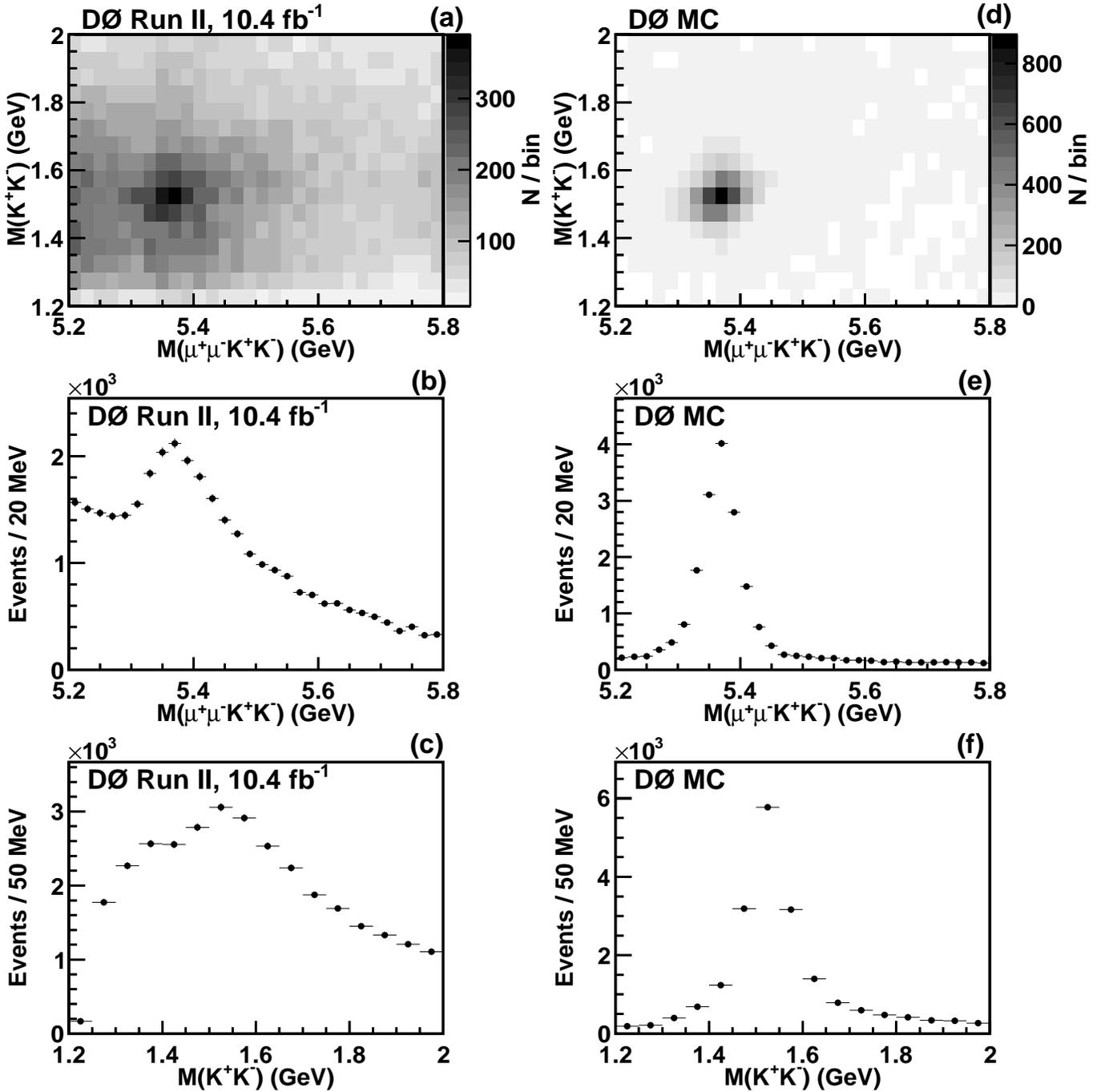}
\caption{(a) Invariant mass of the  $B^0_s$ candidates
as a function of $M(K^+K^-)$ as well as (b) and (c) one-dimensional projections.
The observed  structure is consistent with a decay
$B_s^0 \rightarrow J/\psi K^+K^-$ proceeding through
$f_2^{\prime}(1525)$ or  $f_0(1500)$ mesons. 
(d) Invariant mass of the  $B^0_s$ mesons as a function of $M(K^+K^-)$ 
for the simulated decay \bsf\, as well as (e) and (f) one-dimensional projections.
}
\label{fig:kkvsbs_data}
\end{center}
\end{figure}


\begin{figure}
\begin{center}
\includegraphics*[width=0.42\textwidth]{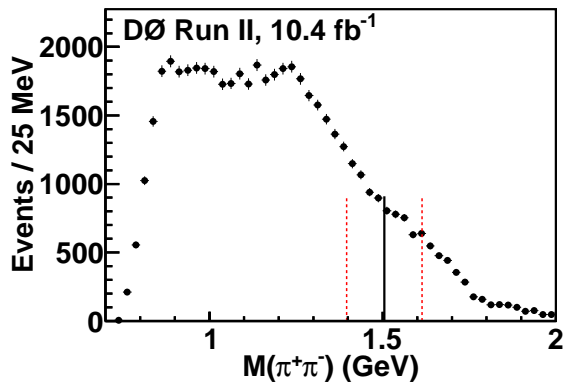}
\caption{Distribution of the  invariant mass $M(\pi^+\pi^-)$. 
 The solid and dashed vertical lines represent the world-average mass and
 the natural width of the $f_0(1500)$ meson~\cite{PDG}, respectively.
}
\label{fig:pipivsbs}
\end{center}
\end{figure}

\end{widetext}

\begin{widetext}

\begin{figure}[H]
\vspace{0.20cm}
\begin{center}
\includegraphics*[height=18.0cm,width=1.0\textwidth]{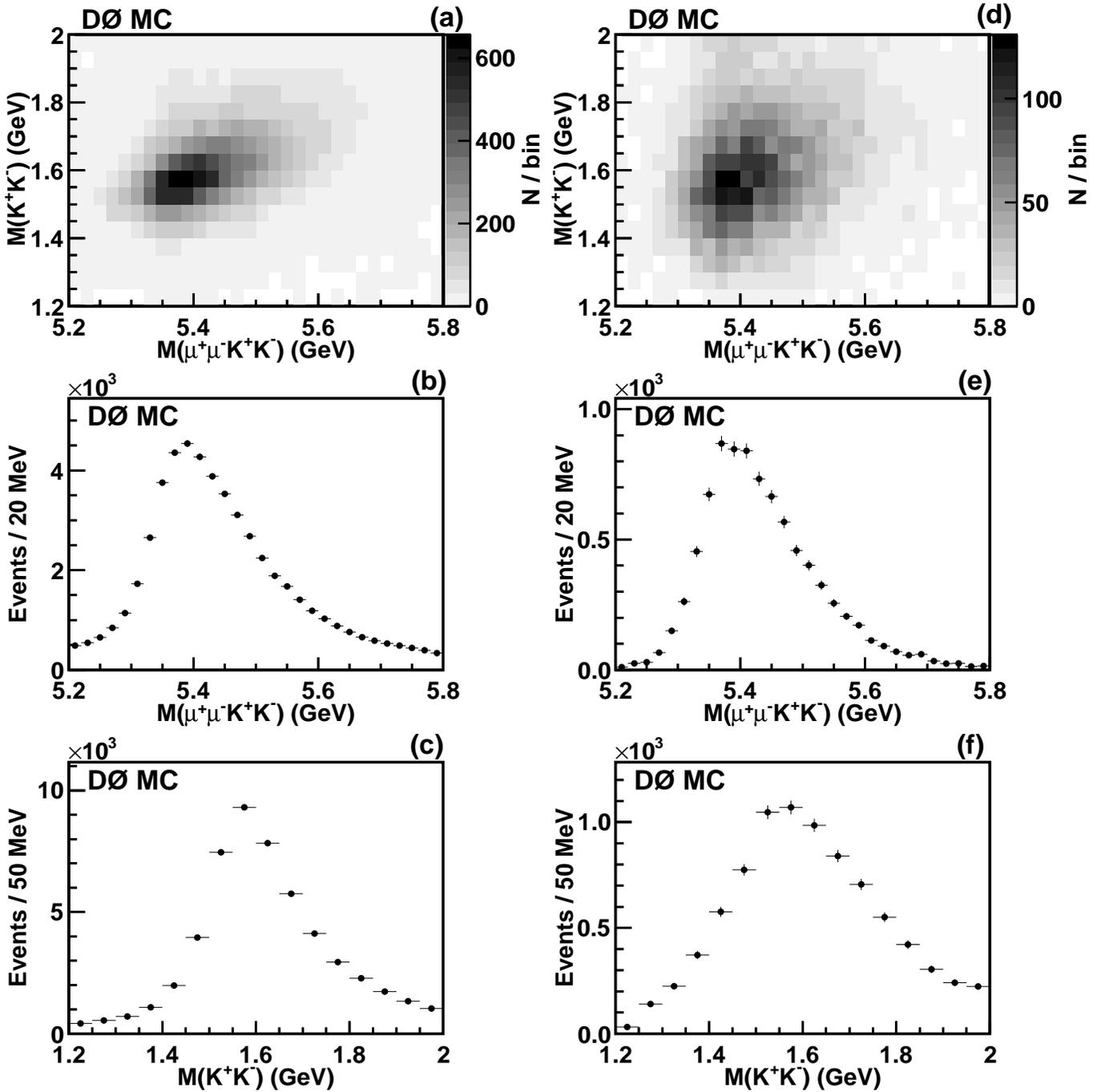}
\caption{(a) Invariant mass of the $B_s^0$ candidates versus $M(K^+K^-)$ for
      simulated decay $B^0 \rightarrow J/\psi K_2^*(1430); K_2^* \rightarrow
      \pi^{\pm} K^{\mp}$, with the pion assigned the kaon mass, as well as
    (b) and (c) one-dimensional projections.
      (d) $K\pi$ $\cal {S}$-wave contribution~\cite{babarz} to the decay
      $B^0 \rightarrow J/\psi K^{\pm}\pi^{\mp}$, with the pion assigned
      the kaon mass as well as (e) and (f) one-dimensional projections.
}
\label{fig:kkvsbsfkst}
\end{center}
\end{figure}

\end{widetext}

 There are two  $K^*_{J}(1430)$ states,
degenerate in mass  but differing in width, $\Gamma = 0.109 \pm
0.005$~GeV for $J=2$ and $\Gamma=0.27 \pm 0.08$~GeV for $J=0$~\cite{PDG}. 
Simulated distributions of $B^0_s$ mass versus  $M(K^+K^-)$  for misidentified decays 
$B^0 \rightarrow J/\psi K^*_{2}(1430)$ and $B^0 \rightarrow J/\psi K^*_{0}(1430)$ 
are shown in Fig.~\ref{fig:kkvsbsfkst}. In the case of $J=0$, we use the full $I=1/2$
$K\pi$ elastic scattering $\cal {S}$-wave amplitude~\cite{babarz} composed of  $K^*_{0}(1430)$
and a nonresonant term.

As seen in  Fig.~\ref{fig:kkvsbsfkst}, the decays 
$B^0 \rightarrow J/\psi K^*_J(1430)$ with $J=0,2$ can mimic the decay \bsf,
with the apparent $B_s^0$ mass peak position increasing with $M(K^+K^-)$. 
We take this peaking background into account by
constructing templates of the  $B_s^0$ mass distribution in steps of
the $K^+K^-$ invariant mass of 50 MeV.
Examples of the  templates are shown in
Figs.~\ref{fig:templates2} and \ref{fig:templates0}. Figure~\ref{fig:tails} shows  
the simulated \bsf\ signal in the same $M(K^+K^-)$ ranges,
with fits to a sum of two Gaussian functions.

Other possible sources of peaking background include B-meson decays
to $J/\psi$ mesons accompanied by the $J=1$ and $J=3$ resonances $K^{*}(1410)$
and $K^{*}_3(1780)$. The former decays predominantly to $K^{*}(892)$.
When the kaon mass is used for both tracks, the latter resonance
would peak at $M(K^+K^-)$  greater than 1.78~GeV.

The candidate mass distribution in the  range $1.45 <M(K^+K^-) <1.60$ GeV 
and $|\cos \psi|<0.8$ is shown in Fig.~\ref{fig:bsmasskksig}, where
$\psi$ is the helicity angle defined in Sec.~\ref{sec:spin}. 
A fit of templates for the \bskk\ decay, the  $B^0 \rightarrow J/\psi K^*_2(1430)$ and
$\cal S$-wave $K\pi$  background
and a linear combinatorial background, yields $534\pm 101$ 
\bskk\ events. The relative rate of the $\cal S$ and $\cal D$ $K\pi$ wave
is constrained to the ratio of 3:2 reported in Ref.~\cite{babarz}.  

To extract the \bs\ signal yield, we use the $B_s^0$  mass
distribution for $M(K^+K^-)$ between 1.35 and 2~GeV.
We fit the simulated signal templates
to the data, with the mass parameter of the core Gaussian function
in each $M(K^+K^-)$ bin scaled by a factor  of $0.9982\pm0.0008$ obtained by matching
the simulation and data as shown in Fig.~\ref{fig:bsmasskksig}.
Figure~\ref{fig:massbins_kst02}  shows the $B_s^0$ mass fits
using the  templates for the signal \bsf\ and for the $B^0 \rightarrow J/\psi K^*_2(1430)$
and  $B^0 \rightarrow J/\psi K^*_0(1430)$
reflections. In these fits we allow the relative rates of the $\cal S$ and $\cal D$ $K\pi$
contributions to vary.
Note that a nonresonant $K^+K^-$  component is implicitly included in the
signal part of these fits.
In addition to the peaking background there is a
background due to random combinations and partially
reconstructed $B$-meson decays, described by a linear function. 
We allow the relative normalization of the
two $K^*_J$ states to vary  in each fit.
The normalization parameters of the \bs\ signal and background components are not 
constrained to be positive  in order to obtain unbiased results for rates close to zero.
We have conducted toy MC ensemble tests and we confirm that there are no biases on signal yield
introduced by the described fitting procedure.

\begin{figure}[]
\begin{center}
\includegraphics*[width=0.45\textwidth]{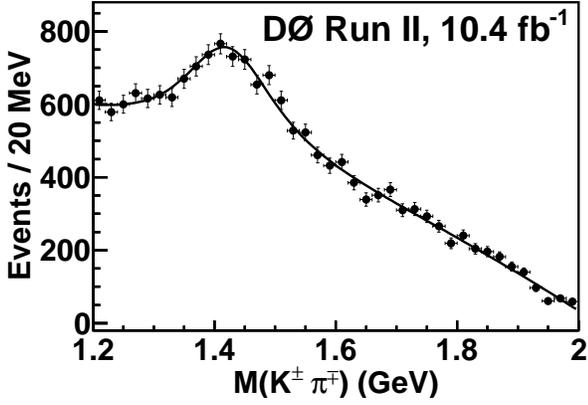}
\caption{Invariant mass distribution of the meson pair from 
the  $B_s^0$ candidates
in the mass range $5.4 <M(J/\psi K^+K^-) <5.6$~GeV, under the 
 $K^{\pm}\pi^{\mp}$ hypothesis. 
The curve shows the  fit of a relativistic Breit-Wigner $J=2$ resonance 
at a fixed mass of 1.43~GeV and with a floating width, over a background
described by a second-order polynomial  function.
}
\label{fig:kpimassfit}
\end{center}
\end{figure}

As seen from Fig.~\ref{fig:massbins_kst02}, 
the fitted yields for the $B^0 \rightarrow J/\psi K^*_2(1430)$ and
$B^0 \rightarrow J/\psi K^*_0(1430)$
decays exceed the $B_s^0$ signal in most of the 11 subsamples,
although the data do not provide a significant
constraint on their relative strength. 
For an independent study of this background, we select events
in the range $5.4< M(J/\psi K^+K^-)<5.6$~GeV, where 
the $B^0 \rightarrow J/\psi K^*_J(1430)$ decays dominate over the $B_s^0$ signal.
In Fig.~\ref{fig:kpimassfit}
we show  the $M(K^{\pm}\pi^{\mp})$ distribution for these events. 
A fit of a relativistic Breit-Wigner $J=2$ resonance 
at a fixed mass of 1.43~GeV and with a floating width, over a background
described by a second-order polynomial  function,
yields  $3386\pm390$ $K^{*}(1430)$ resonance events.
This is in agreement with the total number of events ascribed to the $K^{*}(1430)$
reflection  in this mass range. The best fit result for the width is
$\Gamma = 0.162 \pm 0.019$~GeV that is in between
the widths of the $J=2$ and $J=0$ states.
This study shows that we cannot  establish
the dominance of one background component over the other with the present data.
Figure~\ref{fig:datakkmass_kst02} shows  the
$B_s^0$ signal yield versus $M(K^+K^-)$ from data after taking into account
the $B^0 \rightarrow J/\psi K^*_2(1430)$ and
$B^0 \rightarrow J/\psi K^*_0(1430)$ templates and a linear background.

\section{\label{sec:spin}  Spin of $K^+K^-$ State}
In Section~\ref{sec:signal}, we have presented evidence for the decay
$B^0_s \rightarrow J/\psi K^+K^-$  in the range  $M(K^+K^-)>1.35$~GeV.
The  $M(K^+K^-)$ distribution peaks near 1.5 GeV.
In this section we study the spin of the $K^+K^-$ system 
in the  range $1.45 <M(K^+K^-) <1.60$ GeV. 

A $K^+K^-$ system can be in any natural parity state, $J^P = 0^+, 1^-, 2^+$, etc. 
We consider the values $J = $ 0, 1, and 2 for the spin of the observed structure.

\clearpage

\begin{widetext}

\begin{figure}[]
\begin{center}
\includegraphics*[height=5.5cm,width=1.00\textwidth]{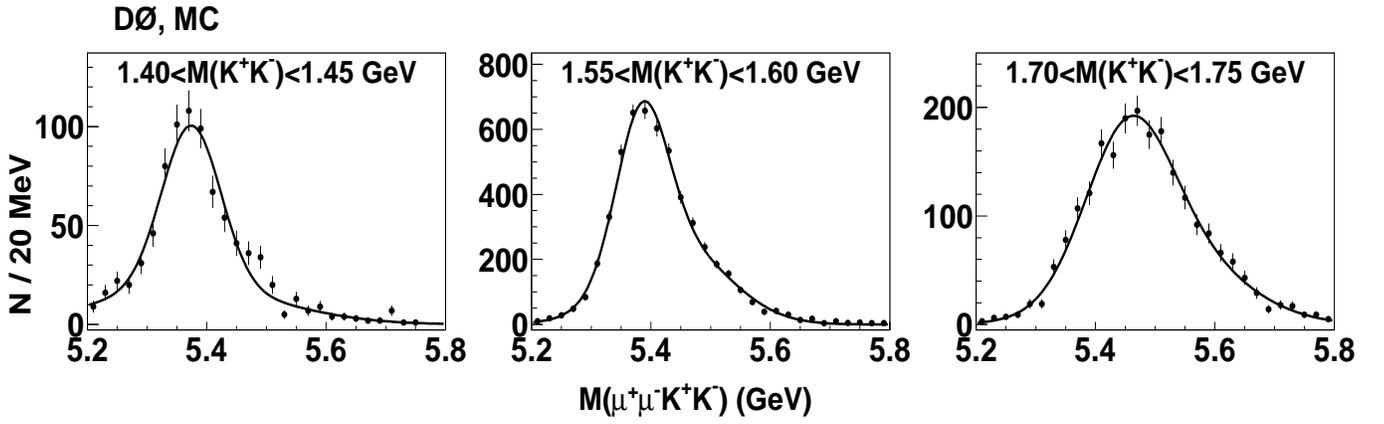}
\caption{Invariant mass of $B^0$ mesons from the simulated 
decay  $B^0 \rightarrow J/\psi K^*_2(1430),  K^*_2(1430) \rightarrow K^{\pm} \pi^{\mp} $,
where the pion is assigned the kaon mass, for a sampling of different $M(K^+K^-)$ ranges.
The distributions are fitted with a sum of two Gaussian functions
with free masses, widths, and normalizations.
}
\label{fig:templates2}
\end{center}
\end{figure}

\begin{figure}[]
\begin{center}
\includegraphics*[height=5.5cm,width=1.00\textwidth]{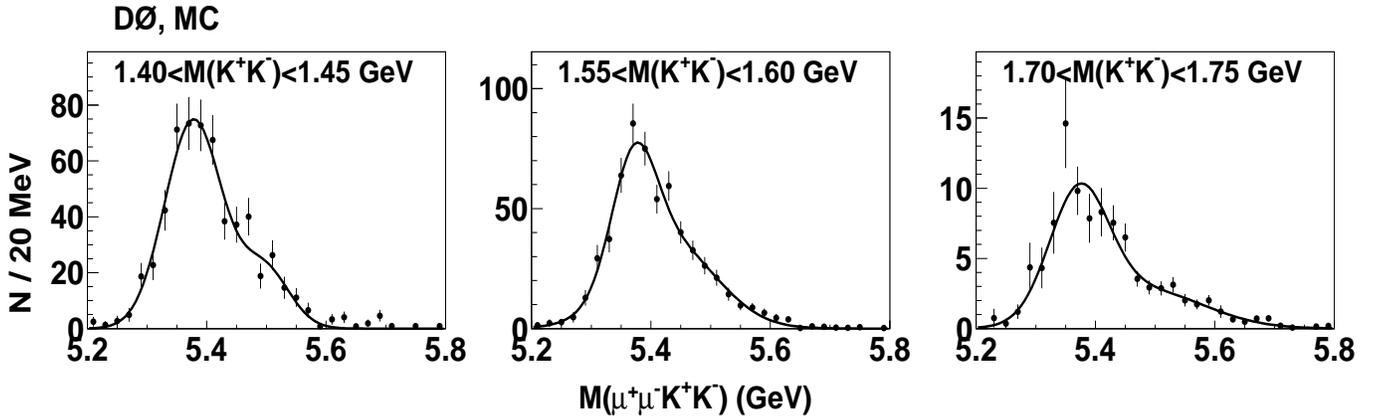}
\caption{Invariant mass of $B^0$ mesons from the simulated
$K\pi$ $\cal {S}$-wave contribution~\cite{babarz} to the decay 
  $B^0 \rightarrow J/\psi K^{\pm}\pi^{\mp}$, 
where the pion is assigned the kaon mass, for a sampling of different $M(K^+K^-)$ ranges.
The distributions are fitted with a sum of two Gaussian functions
with free masses, widths, and relative normalizations.
}
\label{fig:templates0}
\end{center}
\end{figure}

\begin{figure}[]
\begin{center}
\includegraphics*[height=5.5cm,width=1.0\textwidth]{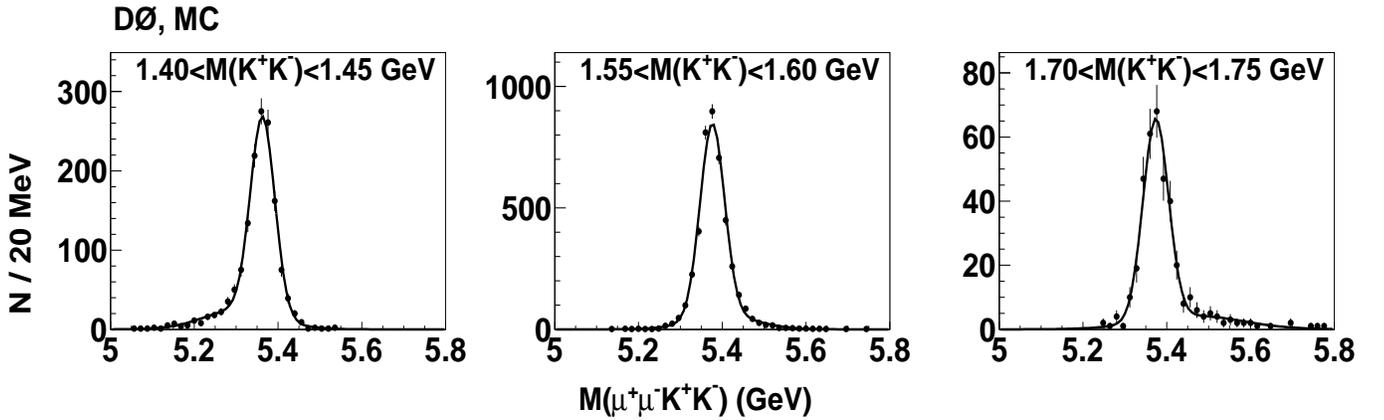}
\caption{ Invariant mass of $B_s^0$ mesons from the simulated decay \bsf\ for a sampling of different ranges of
$M(K^+K^-)$. The distributions are fitted with a sum of two Gaussian functions
with free masses, widths, and relative normalization.
}
\label{fig:tails}
\end{center}
\end{figure}

\clearpage

\begin{figure}[]
\begin{center}
\includegraphics*[height=6.4cm,width=0.99\textwidth]{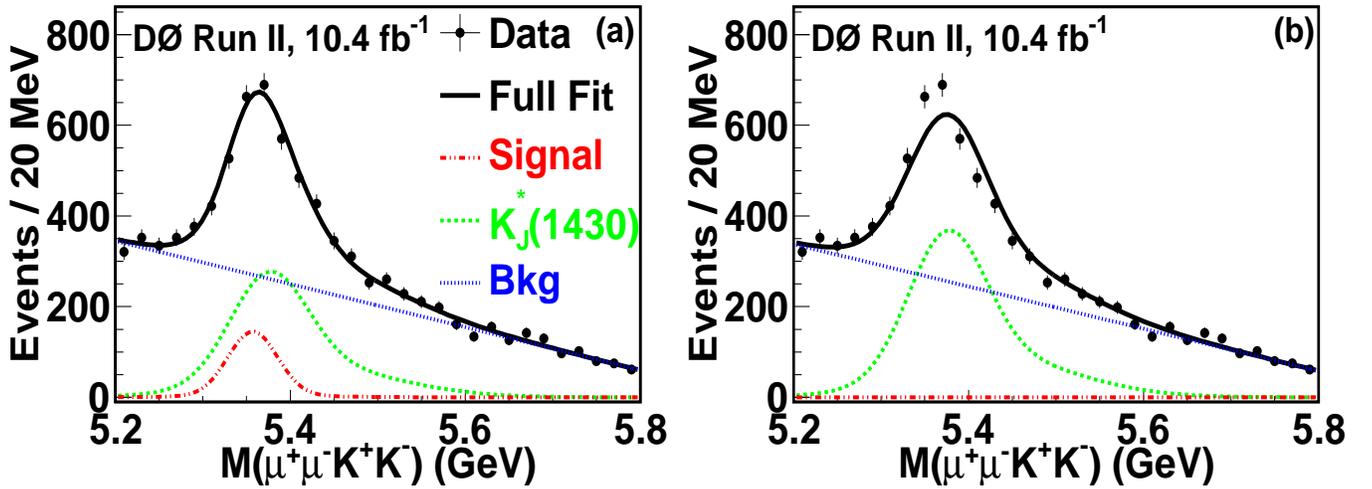}
\caption{ Invariant mass of the $B_s^0$ candidates with 
$1.45 < M(K^+K^-)<1.6$ GeV. Curves show  (a) the result of the fit allowing for  a free
\bs\ signal yield (dashed-dotted lines)  and a background composed of  a 3:2 mixture of 
$B^0 \rightarrow J/\psi K^{\pm} \pi^{\mp}$
decays with the  $K^{\pm} \pi^{\mp}$
system in the $J=0$ and $J=2$ states
(dashed lines) and a combinatorial component described by a linear
function (dotted lines), and
(b) the result of the fit  assuming the same background model but
setting the \bs\ signal rate to zero.
}
\label{fig:bsmasskksig}
\end{center}
\end{figure}

\begin{figure}[]
\begin{center}
\includegraphics*[height=16.0cm,width=0.9\textwidth]{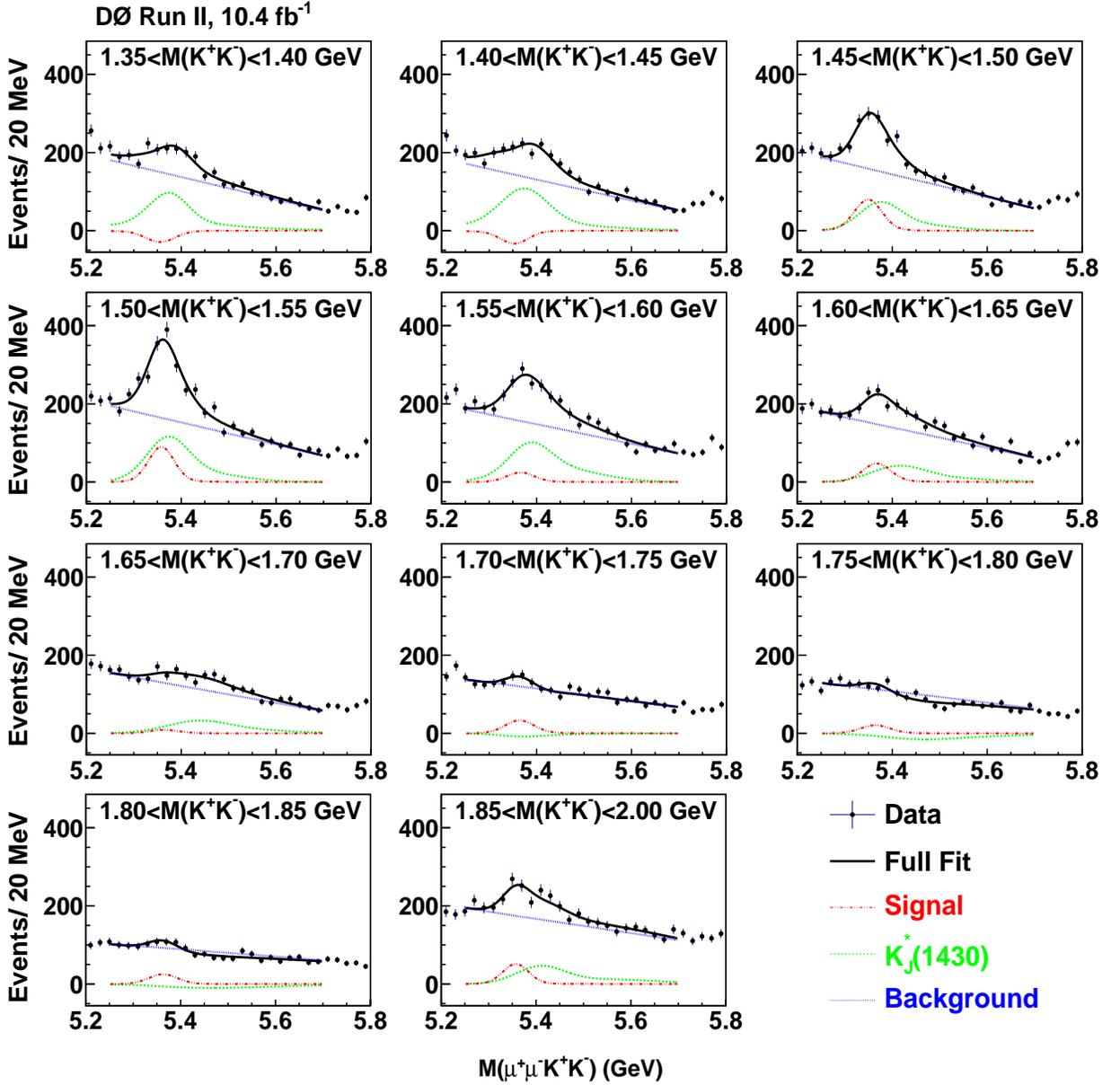}
\caption{Invariant mass of candidates for the decay  \bskk\
in  a sampling of different $M(K^+K^-)$ ranges.
Each fit uses a template derived from the  $B_s^0$ signal simulation, a combination of the 
templates for the
decays  $B^0 \rightarrow J/\psi K^*_J(1430),  K^*_J(1430) \rightarrow K^{\pm} \pi^{\mp}$, $J=1,2$, 
as shown in Figs.~\ref{fig:templates2}  and~\ref{fig:templates0}, and a linear
function  describing the combinatorial background.
The fit is performed in the range $5.25 < M(J/\psi K^+K^-)<5.7$~GeV,  
used to avoid the steeply falling background from multibody decays of $B$ mesons
at lower masses and the steeply rising background from decays $B^{\pm}
\rightarrow J/\psi K^{\pm}$ at higher masses. 
Neither the ~\bs\ signal nor background components are
constrained to positive values in order to obtain unbiased results for rates close to zero.
}
\label{fig:massbins_kst02}
\end{center}
\end{figure}

\end{widetext}

\begin{figure}[]
\begin{center}
\includegraphics*[width=0.45\textwidth]{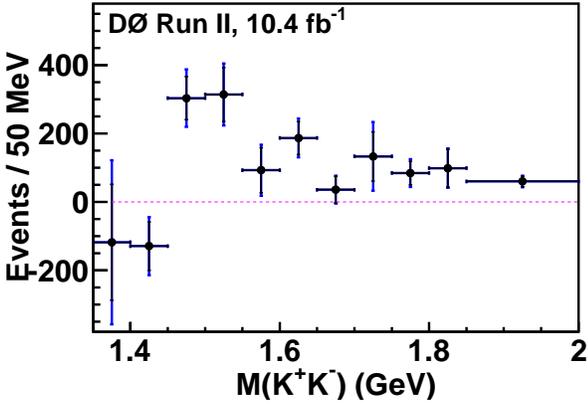}
\caption{The $B_s^0$ signal yield as a function of $M(K^+K^-)$ obtained
from the fits shown in Fig.~\ref{fig:massbins_kst02}. The outer and inner error bars
correspond to the statistical uncertainties with and without systematic uncertainties
added in quadrature.
}
\label{fig:datakkmass_kst02}
\end{center}
\end{figure}

\begin{figure}
\begin{center}
\includegraphics*[width=0.49\textwidth]{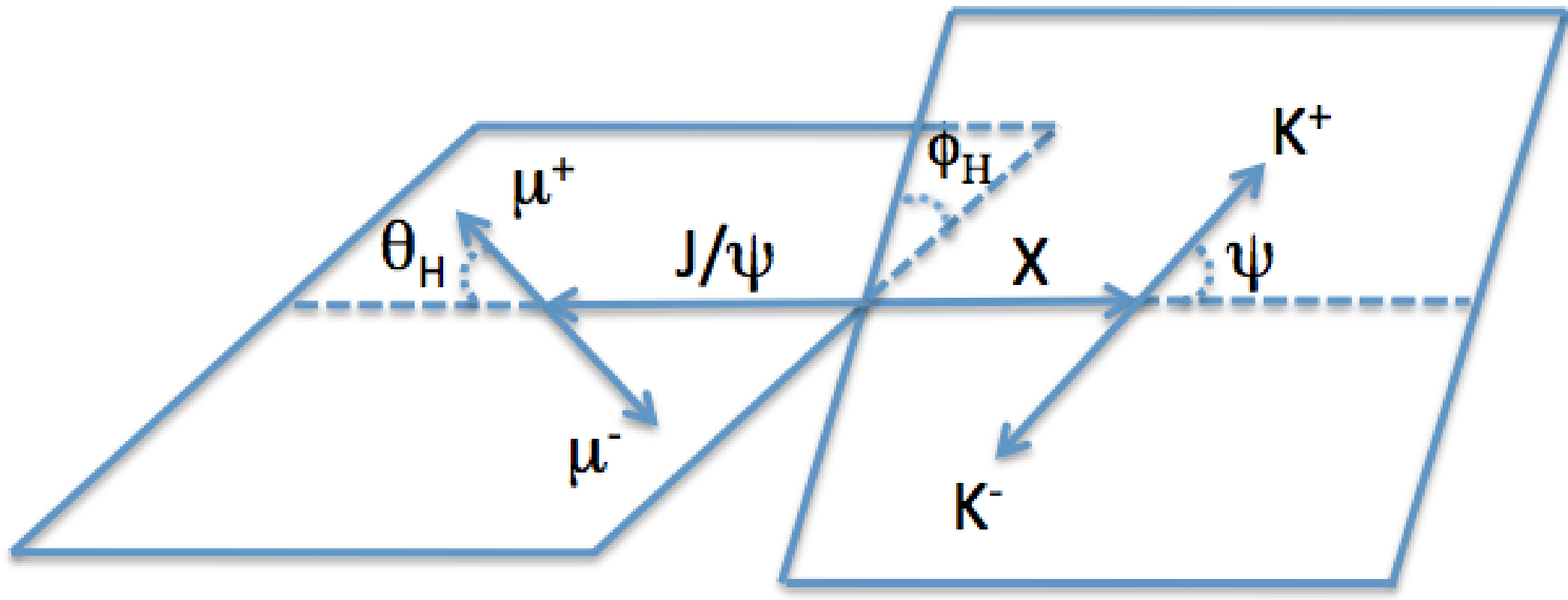}
\caption{Definition of the decay angles $\theta_H$, $\phi_H$, and
$\psi$ in the helicity basis for the sequential decay $B_s^0 \rightarrow
J/\psi X$, $J/\psi \rightarrow \mu^+ \mu^-$,  $X \rightarrow K^+K^-$.
}
\label{fig:helicity}
\end{center}
\end{figure}

\begin{figure}
\begin{center}
\includegraphics*[height=14cm,width=0.35\textwidth]{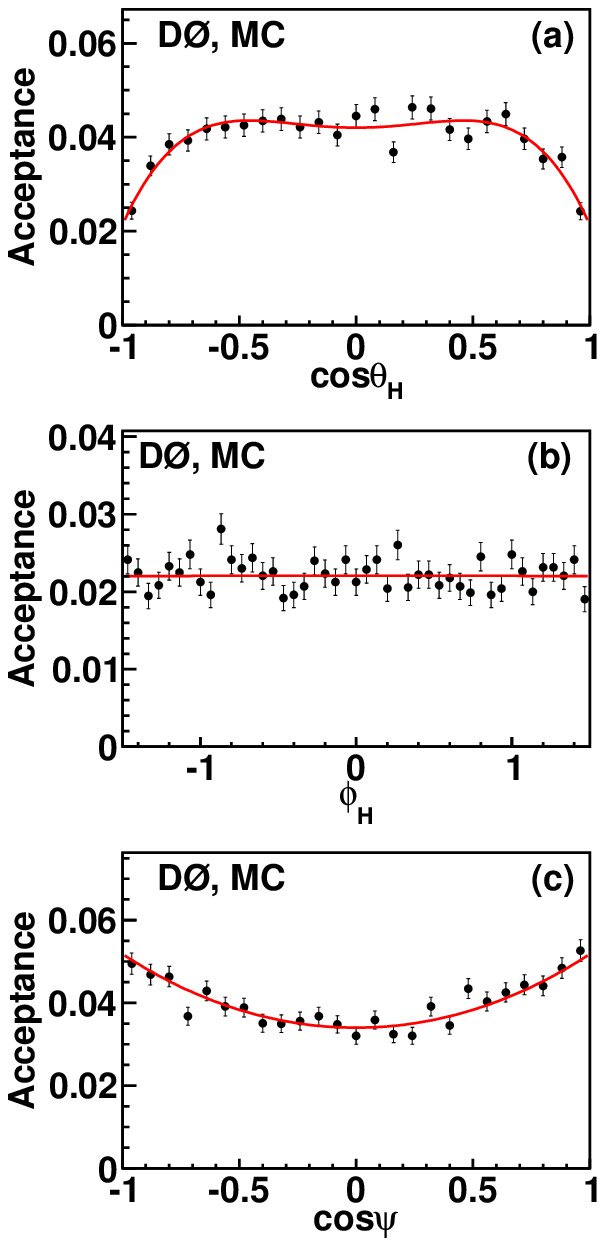}
\caption{Angular dependence of the detection and event selection acceptance of
the decay \bsf\ from simulations as a function of
(a)  $\cos \theta_H$, (b) $\phi_H$,   and (c) $\cos \psi$. The acceptance is found to be
independent of the angle $\phi_H$. For  $\cos \theta_H$ and
$\cos \psi$, we fit the acceptance dependence with symmetric
fourth-order polynomial functions. 
}
\label{fig:angaccept}
\end{center}
\end{figure}

The final state can be described by three independent angles.
We define them as follows: 
$\theta_H$ is the angle between the direction of $\mu^+$ and
$B_s^0$  direction  in the $J/\psi$ rest frame, 
$\psi$ is the angle of the $K^+$ meson
with respect to the $B_s^0$ direction in the $K^+K^-$ rest frame,
and $\phi_H$ is the angle between the two decay planes, as shown in  Fig.~\ref{fig:helicity}.
The angular distribution for the decay of a spinless meson
into the  spin-one meson $J/\psi$   and a meson of unknown spin $J$
can be expressed in terms of $H_1 = \cos\theta_H$, $H_2=\cos\psi$, and $\phi_H$
as follows~\cite{datta}:

\begin{eqnarray}
 \frac {d \Gamma}{ d \Omega}  & \propto & \bigl \lvert \Sigma  A_{Jm} Y_1^{m}(H_1, \phi_H) Y_J^{-m}(-H_2,0) \bigr \rvert^2  D( \Omega),
\end{eqnarray}
where $Y_J^m$ are spherical harmonics, $A_{Jm}$ are complex amplitudes corresponding to
spin $J$  and helicity $m$, and $\Omega$ is either $H_1$, $\phi_H$ or
$H_2$, and the sum extends over equal helicities of the daughter particles,
$m=0$ or $m=\pm1$. The factor $D$ is the acceptance of the event selection. 
Its dependence on the three angular variables is shown in Fig.~\ref{fig:angaccept}.

\begin{figure}[]
\begin{center}
\includegraphics*[height=11.0cm,width=0.48\textwidth]{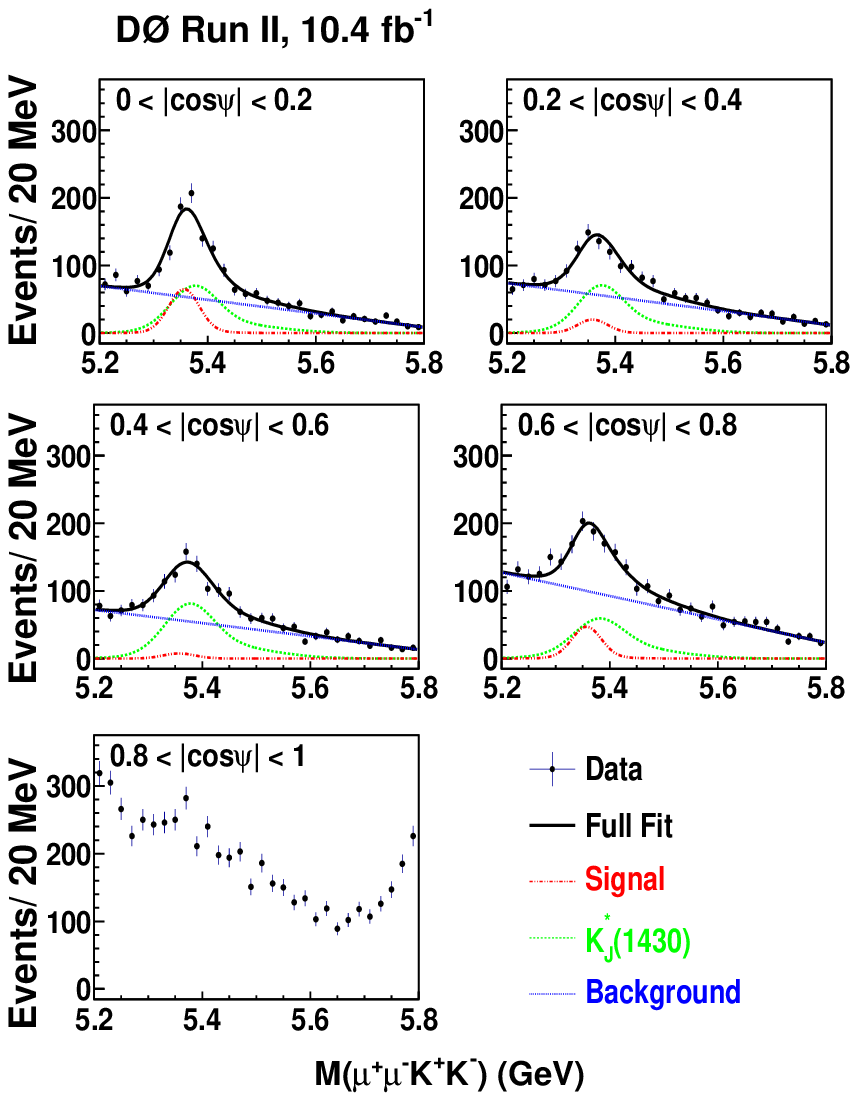}
\caption{The $B_s^0$ mass distribution in five intervals of  $|\cos \psi|$ as shown.
The fits assume a two-Gaussian \bs\ signal template  and a background composed 
of a  reflection of the $B^0 \rightarrow J/\psi K^{\pm} \pi^{\mp} $ decay  and a linear component.
There is a steeply falling background from multibody decays of
$B$ mesons at lower masses and a steeply rising background from
decays $B^{\pm} \rightarrow J/\psi K^{\pm}$ at higher masses.
These backgrounds are particularly acute for cos$(\psi)>$~0.8
making it impossible to obtain a reliable fit in this region.
}
\label{fig:masscos}
\end{center}
\end{figure}

\begin{figure}
\begin{center}
\includegraphics*[width=0.45\textwidth]{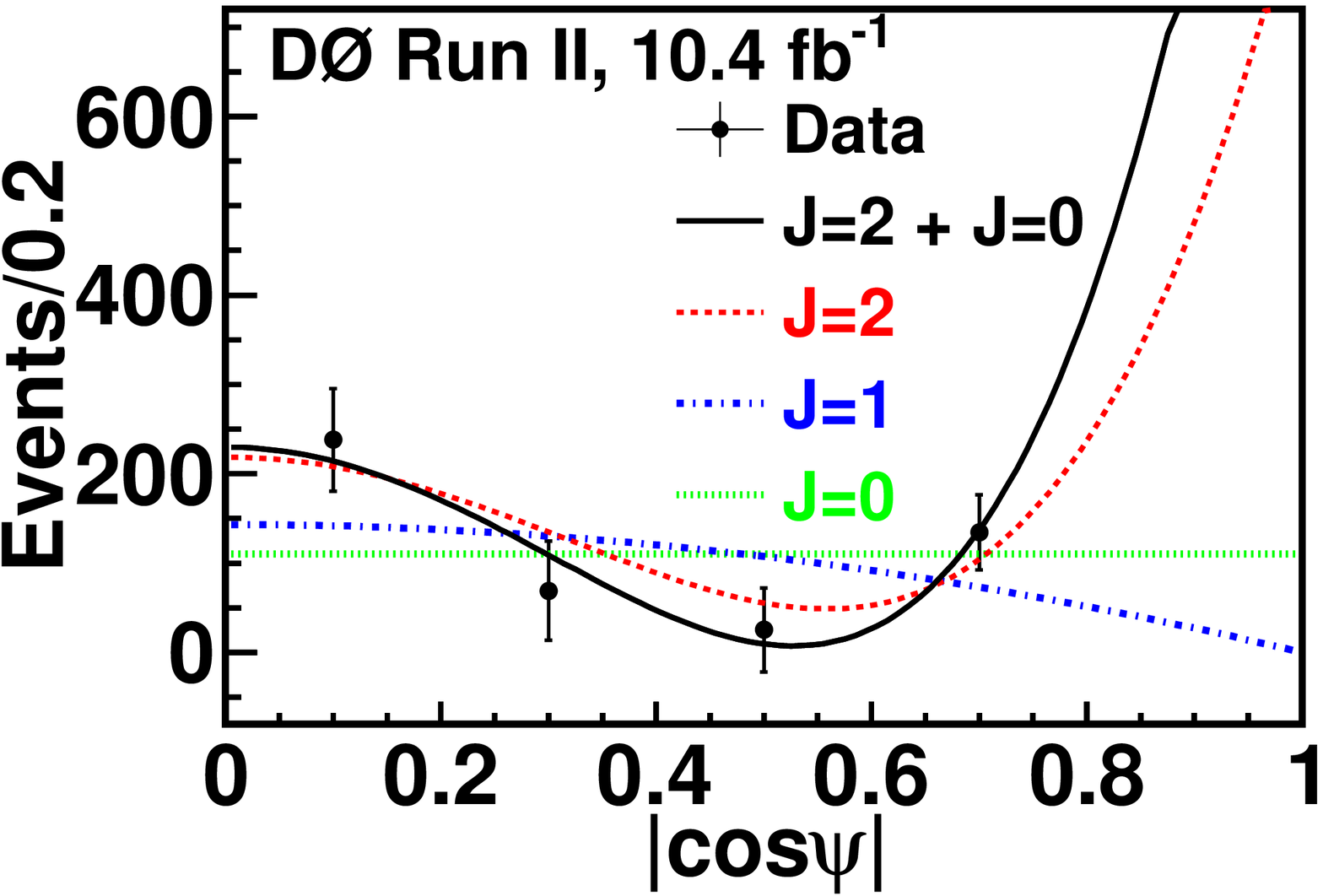}
\caption{The $|\cos \psi|$ distribution for the decay $B_s^0 \rightarrow J/\psi X$,
$X\rightarrow K^+K^-$ in the range $1.45 < M(K^+K^-)<1.6$ GeV.
The curves are best fits assuming  pure $J=0$ (dashed line),
pure $J=1$ (dashed-dotted line), and pure $J=2$ (solid line).
}
\label{fig:spin}
\end{center}
\end{figure}

Due to limited statistics and a large background, we focus on the 
$\cos \psi$ distribution obtained by integrating the  angular distribution
over  $\cos \theta_H$ and $\phi_H$, taking into account the variation of
the acceptance as a function of  $\cos \theta_H$. We extract the 
\bs\ signal rate as a function of $|\cos \psi|$ by fitting 
the candidate mass in five regions of  $|\cos \psi|$.
The data and fit results are shown in Fig.~\ref{fig:masscos}. 
The resulting distribution, corrected for acceptance, is shown in  
Fig.~\ref{fig:spin}. Systematic uncertainties due to the
shape of  combinatorial background, signal model, and acceptance,
are added in quadrature.
In the region  $|\cos \psi|>0.8$,
the large background prevents obtaining a reliable fit.

For  $J=0$, the expected distribution is isotropic.
For $J=1$, the   $\cos \psi$ distribution without the acceptance factor is given by:
\begin{eqnarray}
\frac {d \Gamma } { d\cos\psi}  & \propto &   F_{10}(2\cos^2\psi)  + (1-F_{10})\sin^2 \psi,
\end{eqnarray}
where $F_{10}$ is the ratio of the rate $J=1, m=0$ to the total $J=1$ rate. 
For a superposition of $J=0$ and 2,
with a free relative normalization, the  $\cos \psi$ distribution is obtained from

\begin{eqnarray}
\lefteqn{ \frac {d \Gamma } { d\cos\psi} \propto } \nonumber \\
& &  \Bigl \lvert \sqrt{F_{20}(1-F_0)(5/4)} (3\cos^2\psi-1)  + \exp(i \delta_{0}) \sqrt{ F_{0} } \Bigr \rvert ^2 \nonumber \\
& & + \frac {15}{2} (1-F_{20})(1-F_0)\sin^2\psi (1 - \sin^2 \psi),
\end{eqnarray}
where $F_{20}$ is the ratio of the rate $J=2$, $m=0$ to the total 
$J=2$ rate, and $F_0$ is the $J=0$ fraction with relative phase angle $\delta_{0}$.

Figure~\ref{fig:spin} shows that the data favor $J=2$, hence the peak is identified with the $f_2^{\prime}(1525)$ meson.
The fit probabilities for pure $J=0$ and pure $J=1$ are
$2.8 \times 10^{-2}$  and $9.8 \times 10^{-3}$,  respectively. 
For  $J=2$ the fit probability is 0.27.
The data are also consistent with  a coherent superposition of
$J=0$ and $J=2$ states.  
With $F_{20}=1$, we obtain the $\cal {S}$-wave fraction of  $F_0 = 0.06 \pm 0.16$
and a fit probability of 0.37.

\section{\label{sec:sigyield}Signal Yield}

The measured decay rate of a particle resonance 
as a function of the invariant mass  of the final state, is described by
the relativistic Breit-Wigner function (RBW)~\cite{PDG} 
convoluted with  detector resolution.

\begin{figure}[htbp]
\begin{center}
\includegraphics*[width=0.45\textwidth]{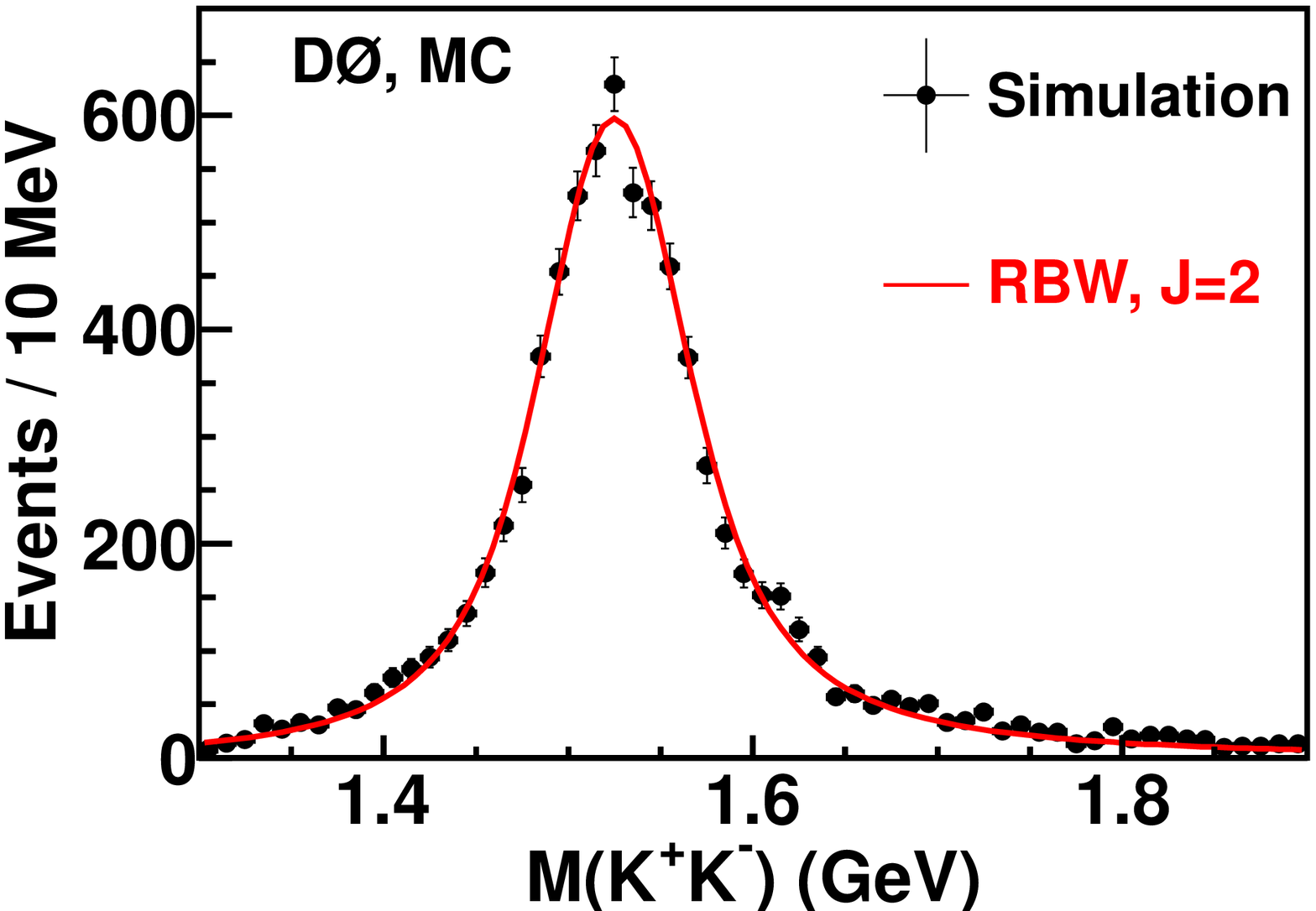}
\caption{The $M(K^+K^-)$  distribution from the simulation of the decay \bsf\
fitted by the relativistic Breit-Wigner function~\cite{PDG}.}
\label{fig:fitmkkmc}
\end{center}
\end{figure}

To obtain the detector resolution, we use simulated 
\bsf\ decays where the $J/\psi$ is forced to decay into two muons and
the $f_2^{\prime}(1525)$ into two kaons. The fitted $M(K^+K^-)$ 
distribution for the simulation is shown in Fig.~\ref{fig:fitmkkmc}. 
Fixing the mass and natural width parameters at their input values,
$M = 1525$ MeV and $\Gamma_0 = 73$ MeV, and using the
range parameter~\cite{PDG} $R=5.0$~GeV$^{-1}$, 
we obtain $\sigma(M) = 22 \pm 1$ MeV.

\begin{figure}[htbp]
\begin{center}
\includegraphics*[width=0.45\textwidth]{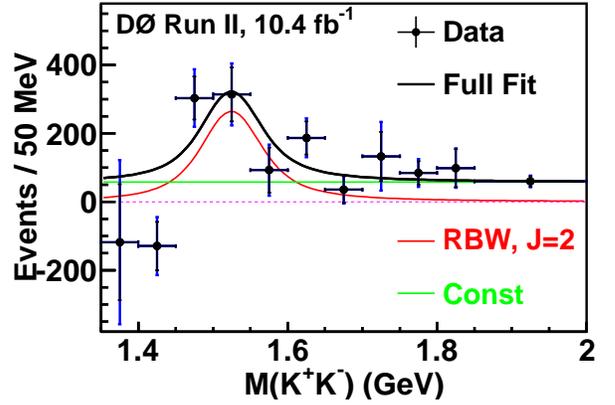}
\caption{ Fit to the $B_s^0$ yield  versus  $M(K^+K^-)$ as obtained in
  Fig.~\ref{fig:datakkmass_kst02}. The full fit to $B_s^0 \rightarrow  J/\psi K^+K^-$ includes a
$f_2^{\prime}(1525)$ signal described by a relativistic Breit-Wigner and
nonresonant constant term for the $\cal S$ wave. 
The outer and inner error bars correspond to the statistical uncertainties with
and without systematic uncertainties added in quadrature.
}
\label{fig:datakkmassfit_kst02}
\end{center}
\end{figure}

We fit the $B_s^0$ signal yield versus $M(K^+K^-)$ from data, as shown in
Fig.~\ref{fig:datakkmass_kst02}, to an incoherent  sum of the $J=2$ component 
 and a constant continuum term. The result is
shown in Fig.~\ref{fig:datakkmassfit_kst02}. 
The fit yields 629$\pm$157   $f_2^{\prime}(1525)$ events and 345$\pm$76 events 
for  the constant term in the mass range  $1.4  <M(K^+K^-)<1.7$~GeV.
The  fraction of the nonresonant term, assumed to be the $\cal S$
wave, in this mass range is $0.35\pm0.09$.

\section{\label{sec:BFratio} Ratio of  ${\cal B} (B_s^{0} \rightarrow J/\psi f_{2}^{\prime}(1525))$ and ${\cal B} (B_s^0 \rightarrow J/\psi \phi)$ }

To determine an absolute branching fraction for the \bsf\ decay, efficiencies, branching fractions,
and the cross section need to be known, as well as the integrated luminosity. 
However, terms common to  the $B_{s}^{0}\rightarrow J/\psi f_{2}^{\prime}(1525)$ 
and the $B_{s}^{0}\rightarrow J/\psi \phi$ branching fractions cancel in their ratio.
A measurement of  the  relative branching fraction $R_{f_2^{\prime}/\phi}$ requires
the yields of the two decays, 
$N_{ B_{s}^{0} \rightarrow J/\psi f_{2}^{\prime}(1525)}$ and $N_{B_{s}^{0} \rightarrow J/\psi \phi}$,
and the reconstruction efficiencies of
the two decay modes,  $\varepsilon_\mathrm{reco}^{B_{s}^{0} \rightarrow J/\psi \phi}$ 
and $\varepsilon_\mathrm{reco}^{B_{s}^{0} \rightarrow J/\psi f_{2}^{\prime}(1525)}$:
\begin{eqnarray}
R_{f_2^{\prime}/\phi} & = & {\frac{ \mathcal{B} (B_{s}^{0}\rightarrow J/\psi f_{2}^{\prime}(1525)    ;f_{2}^{\prime}(1525)    \rightarrow K^{+} K^{-})  }
       { \mathcal{B} (B_{s}^{0}\rightarrow J/\psi \phi;\phi \rightarrow K^{+} K^{-}    )}}  \nonumber \\
 &=& \frac{N_{ B_{s}^{0} \rightarrow J/\psi f_{2}^{\prime}(1525)} \times \varepsilon_\mathrm{reco}^{B_{s}^{0} \rightarrow J/\psi \phi}}
{N_{B_{s}^{0} \rightarrow J/\psi \phi} \times \varepsilon_\mathrm{reco}^{B_{s}^{0} \rightarrow J/\psi f_{2}^{\prime}(1525)} },
\end{eqnarray}
where $N_{ B_{s}^{0} \rightarrow J/\psi f_{2}^{\prime}(1525)}$ is  determined
in the mass range  $1.4 < M(K^+K^-) <1.7$~GeV and
$N_{B_{s}^{0} \rightarrow J/\psi \phi}$ in the mass range
 $1.01 < M(K^+K^-) <1.03$~GeV. The only
difference in the event selection for the two channels is the $M(K\pi)>1$ ~GeV
 condition  applied for the  $J/\psi f_{2}^{\prime}(1525)$  candidates.

\begin{figure}[h]
\begin{center}
\includegraphics*[width=0.45\textwidth]{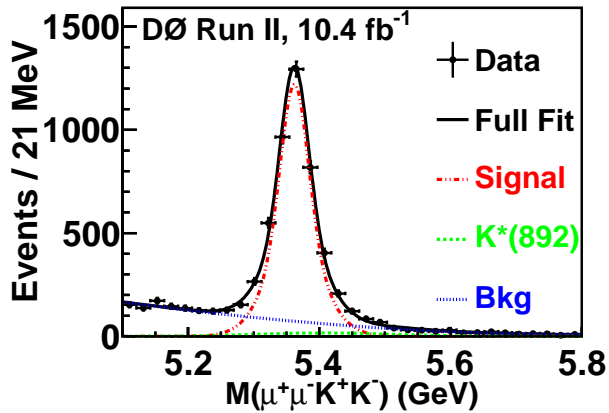}
\caption{Invariant mass distribution of 
 \bs\  candidates with $ct>200$ $\mu$m for events in the mass range  $1.01<M(K^+K^-)<1.03$~GeV.
A fit  to a sum of a double Gaussian \bsdec\ signal (dashed-dotted
line),  a quadratic combinatorial background (dotted line),
and the reflection of the decay $B^0 \rightarrow J/\psi K^*(892)$
(dashed line) is used to extract
the \bs\ yield.
}
\label{fig:psiphimass}
\end{center}
\end{figure}

The yield of the $B_{s}^{0}\rightarrow J/\psi \phi$ decay is determined by fitting the data, shown in
Fig.~\ref{fig:psiphimass}, with
a double Gaussian function for the signal, a second-order  polynomial for background,  
and the reflection of the decay \bddec\ taken from simulations.
The total number of  \bsdec\ events is 4064 $\pm$ 105.

We use simulated samples of the two decay processes to determine the reconstruction efficiencies.
For the decay $B_{s}^{0} \rightarrow J/\psi f_{2}^{\prime}(1525)$ 
the efficiency is measured to be $(0.122 \pm 0.002)$\% and
for the decay \bsdec\ it is
$(0.149 \pm 0.002)$\% (where the uncertainties are due to MC
statistics), yielding
$R_{f_2^{\prime}/\phi} = 0.19 \pm 0.05 \rm \thinspace{(stat)}$.

The denominator in Eq.~4 may include a contribution from
the $K^+K^-$ $\cal S$ wave, and no correction is made, allowing the ratio
to be recalculated for different $\cal S$-wave fraction inputs.

\section{\label{sec:syst} Systematic uncertainties}

The main contributions to systematic uncertainties are summarized in Table \ref{syst}.
They are evaluated as follows:

\begin{itemize}

\item {\it $K^*_0(1430)$ width:}
We vary the  $K^*_0$ width within its uncertainty of
0.08~GeV~\cite{PDG}.

\item {\it $K^*_0(1430)$ and $K^*_2(1430)$ templates:}
We vary the shape of the $K^*_0$ and $K^*_2$ templates by altering the
widths of the dominant Gaussian component within statistical uncertainties.

\item {\it Combinatorial background shape:}
As an alternative, we use a second-order polynomial to describe
 the combinatorial background. We also vary the fitting mass range from $5.25 - 5.70$ GeV to $5.2 - 5.8$ GeV.

\item {\it Signal shape:}
We vary the $B_s^0$ mass scale within its uncertainty in data of 0.08\%
and the width of the core Gaussian component by $\pm$10\%.

\item {\it Trigger efficiency:}
Due to the mass difference between the $f_{2}^{\prime}(1525)$ and $\phi$ resonances,
there is a small difference between average muon momenta in the two channels.
Approximately 3\% more $J/\psi \phi$ events have
a leading muon with $p_T>15$~GeV and  about 3\% more $J/\psi \phi$  events have
both muons with $p_T>3$~GeV. We therefore  estimate that there is approximately
a 3\% difference in the fraction of events that can be accepted by the
trigger between the $J/\psi \phi$ and $J/\psi f_2^{\prime}(1520)$ signals.
Trigger simulations confirm this estimate.
We apply the 3\% correction to $R_{f_2^{\prime}/\phi}$ and 
assign an absolute 3\% systematic uncertainty.

\item {\it $M(K^+K^-)$ dependence of  efficiency:} 
The $M(K^+K^-)$ dependence of the efficiency for reconstructing the $f_2^{\prime}(1520)$ resonance is obtained
from a simulation. We assign a 2\% uncertainty due to the statistical
precision of the MC sample.

\item {\it Helicity dependence of  efficiency:} 
The \bsdec\ signal acceptance is obtained from a MC sample generated under the
assumption that the final state is not polarized, i.e., with
the final state distributed uniformly in helicity angle $\cos \theta_H$. 
We compare this signal acceptance with distributions corresponding to
pure helicity 0 and 1 and assign a systematic uncertainty equal to the difference.

\item {\it $f_2^{\prime}(1525)$ mass and natural width:}
The uncertainty on the mass of the $f_2^{\prime}(1525)$ resonance 
of 5 MeV~\cite{PDG} leads to the uncertainty on  $R_{f_2^{\prime}/\phi}$
of 3\%, while the uncertainty on the natural width of 6 MeV
leads to the uncertainty on  $R_{f_2^{\prime}/\phi}$ of 0.7\%.

\item {\it ${B_s^0 \rightarrow J/\psi \phi}$ signal shape:}
The $B_s^0 \rightarrow J/\psi \phi$ signal yield is sensitive to
the signal mass model because of the presence of
the $B_s^0 \rightarrow J/\psi K^*(890)$ reflection that peaks near
the signal. We assign the systematic uncertainty
as one half of the difference between fit results for single
and double Gaussian distributions for the signal mass model.
\end{itemize}

\begin{table}[h!tb]
\caption {Sources of  systematic relative uncertainty on $R_{f_2^{\prime}/\phi}$.}
\begin{tabular}{cc}
\hline
Source & Uncertainty (\%) \tabularnewline
\hline
$K^*_0(1430)$ width & 5 \tabularnewline
$K^*_0(1430)$ and $K^*_2(1430)$ templates & 10\tabularnewline
Combinatorial background shape & 10\tabularnewline
Signal shape & 12 \tabularnewline
Trigger efficiency  & 3 \tabularnewline
$M(K^+K^-)$ dependence of efficiency  & 2 \tabularnewline
Helicity dependence of efficiency  & 3 \tabularnewline
$f_2^{\prime}(1525)$ mass  & 3 \tabularnewline
$f_2^{\prime}(1525)$ natural width  & 1 \tabularnewline
${B_s^0 \rightarrow J/\psi \phi}$ signal shape   & 4 \tabularnewline
\hline
Total                      &  20 \tabularnewline
\hline
\hline
\end{tabular}
\label{syst}
\end{table}

\section{\label{sec:conclusions}Summary and Discussion}
\vspace{-0.2cm}
We confirm the observation of the decay
\bsf\ previously observed by the LHCb Collaboration~\cite{lhcbf2} and
measure the ratio of branching fractions of the decays \bsf\ and \bsdec\ to be
$R_{f_2^{\prime}/\phi} = 0.19 \pm 0.05 \rm \thinspace{(stat)} \pm 0.04 \rm \thinspace{(syst)}$.
The fit to the background-subtracted signal
$B_s^0 \rightarrow J/\psi K^+K^-$, assuming an incoherent
sum of the $J=2$ resonance $f_2'(1525)$ and a constant continuum term,
assigns the fraction of $0.35 \pm 0.09$ to the constant
term. The fit to the helicity angle $\psi$ in the $K^+K^-$ rest frame finds
the $K^+K^-$ resonance to be consistent with $J=2$ with a fit
probability of 0.27 and preferred over $J=0$ or $J=1$, for
which the fit probabilities are $2.8 \times 10^{-2}$ 
and $9.8 \times 10^{-3},$ respectively.


%
We thank the staffs at Fermilab and collaborating institutions,
and acknowledge support from the
DOE and NSF (U.S.);
CEA and CNRS/IN2P3 (France);
MON, Rosatom, and RFBR (Russia);
CNPq, FAPERJ, FAPESP, and FUNDUNESP (Brazil);
DAE and DST (India);
Colciencias (Colombia);
CONACyT (Mexico);
NRF (Korea);
FOM (The Netherlands);
STFC and the Royal Society (U.K.);
MSMT and GACR (Czech Republic);
BMBF and DFG (Germany);
SFI (Ireland);
The Swedish Research Council (Sweden);
and
CAS and CNSF (China).
%

\end{document}